\documentclass[times, 10pt,twocolumn]{article}
\usepackage{latex8}
\usepackage{times}
\usepackage{graphicx}
\usepackage{url}
\usepackage{fancybox}
\usepackage[tight]{subfigure}
\usepackage{algorithm}
\usepackage{algorithmic}
\usepackage{verbatim}
\usepackage{epsfig}
\usepackage{calc}
\usepackage[tbtags]{amsmath}
\usepackage{amsfonts}
\usepackage{amssymb}
\usepackage{amsthm}
\newcommand{\Alice}{Alice}
\begin{document}

\title{Anonymous Gossiping}
\author{Anwitaman Datta\\NTU Singapore}
\renewcommand{\baselinestretch}{1}
\maketitle
\thispagestyle{empty}

\begin{abstract}
In this paper we introduce a novel gossiping primitive to support privacy preserving data analytics (PPDA). In contrast to existing computational PPDA primitives such as secure multiparty computation and data randomization based approaches, the proposed primitive ``anonymous gossiping'' is a communication primitive for privacy preserving personalized information aggregation complementing such traditional computational analytics. We realize this novel primitive by composing existing gossiping mechanisms for peer sampling \& information aggregation and onion routing technique for establishing anonymous communication. This is more an `ideas' paper, rather than providing concrete and quantified results.
\end{abstract}

\footnotesize \emph{``It is perfectly monstrous the way people go about nowadays saying things against one, behind one's back, that are absolutely and entirely true.'' --- Oscar Wilde} \normalsize

\section{Introduction}
Information aggregation and mining is often used to obtain collective intelligence and generate a panoramic (macroscopic) view of a system or to device useful recommendation mechanisms. An interesting niche which has been studied for the last decade is that of privacy preserving data mining (PPDM). The essential idea, to quote the seminal paper on PPDM \cite{PPDM}, is ``\emph{Since the primary task in data mining is the development of models about aggregated data, can we develop accurate models without access to precise information in individual data records?}''. The early works on PPDM were based on random perturbation of information. Since then, a new class of PPDM based on secure multiparty computation \cite{PPDMSMC} has also evolved. The trade-offs between the two approaches are mainly on accuracy and computational complexity \& scalability. Research on both these individual families of privacy preserving data analytics (PPDA) as well as hybridized solutions continue in full steam. Privacy preserving data mining in P2P environments \cite{Bhaduri} has also gained considerable attention in recent years.

In this paper we address an orthogonal question. \emph{Can we facilitate collaborative data analytics among users without disclosing the identity of who are participating and contributing the data?} Computational PPDAs do not provide anonymity. Such privacy is important, say, when the analytics is carried out for a specific subset of users with some shared characteristics, such that besides the privacy of the individual records, the users may be interested to even preserve privacy in terms of them having those characteristics.

We propose a communication primitive (anonymous gossiping\footnote{Epidemic information dissemination leveraging anonymous interactions in mobile ad hoc network has been studied in the past, and uses the same name \cite{Birman}, but what we do is completely unrelated.}) which facilitates such privacy. Specifically, we adapt a well studied point-to-point anonymized communication technique, onion routing \cite{onion,Tor} to achieve it. Note that the actual data analytics itself however may be additionally with or without preserving privacy of the individual records. For example, for an anonymized paper submission, it is ok that the reviewers can read the content of the paper, as long as they do not know who wrote the paper. It is in this sense that our mechanism compliments the existing computational primitives. Anonymous gossiping finds ready usage in emerging P2P applications such as user affinity based personalized decentralized search \cite{Gossple,BenderCKMNPSW08} \& personalized recommendation in decentralized online social networks \cite{buchegger}.

While anonymous communication in p2p systems is an old and well studied problem, e.g., Freenet \cite{Freenet}, the novelties of this paper are (i) defining a novel communication primitive for PPDA, and (ii) proposing a concrete way to do so by composing well studied existing building blocks.

In this short paper, we limit ourselves to defining this new problem and sketching a first solution for the same. A more rigorous analysis and evaluation of the proposed mechanism's security, performance, overheads and subsequent necessary optimization or exploration of alternatives are all issues for future study.

In Section \ref{sec:probstatement} we provide a more succinct description of the problem along with a sketch
 of the solution. We elaborate in detail our assumptions and notations in Section \ref{sec:assumptions}
 before providing the anonymous gossiping protocol in Section \ref{sec:anongoss}. We wrap up in Section \ref{sec:References} with concluding remarks highlighting several interesting extensions and research problems that present themselves from the current work.
\section{Problem statement \& solution sketch}
\label{sec:probstatement}
We want to facilitate the following:
\begin{enumerate}
  \item Allow user specific information (lets call it \emph{user profile}) to be used to carry out any kind of personalized aggregation/clustering, etc. Such mechanisms can then be used for various personalized services such as recommendation or query expansion \cite{BenderCKMNPSW08,Gossple}.
  \item Ensure that an user can not be associated with her\footnote{For simplicity, we choose to use the feminine form to address the users, instead of using his/her/its on every occasion.} profile by others, even while individual users benefit from personalization facilitated by analysis of information aggregated from other similar users.
\end{enumerate}

The basic \emph{outline} of a potential solution to achieve the above mentioned objectives comprise of the following steps and building blocks:
\begin{enumerate}
  \item Aggregation task delegation: Each user delegates the task of personalized aggregation to a proxy peer.
  \item Proxy peers interact among each other to carry out the aggregation task on behalf of the users.
  \item Ensure that the proxy peer is oblivious of the identity of the user(s) on whose behalf it carries out aggregation task. This in turn will ensure users' privacy. This necessitates a mechanism for users to assign the task to a proxy without being identified, and also a mechanism for the proxy to still be able to deliver back the aggregated information to the original user without knowing who it is.
\end{enumerate}

Here we describe a mechanism (anonymous gossiping) to achieve the last point. How the aggregation task itself is carried out among the proxies is an orthogonal issue. This includes the issues of both how proxies interact among each other, and how they carry out the data analytics. Anonymous gossiping is generic in that it can be applied to provide user privacy while using arbitrary gossiping algorithms for information aggregation.

Whether any peer is adequate to act as a proxy, or whether some other considerations such as trustworthiness, or betweenness in social graph (facilitating quicker aggregation) etc. need to be taken into account while delegating the task is ignored in the current work.

\section{Assumptions and notations}
\label{sec:assumptions}
Our solution relies on the following assumptions and existing primitives.
\begin{enumerate}
  \item Users form and participate in an overlay. This overlay may be a classical unstructured network or a semantic or social overlay.
  \item Users use public key as their logical identifier in the system. However, there is no need for a public key infrastructure (PKI) since we are not trying to establish if a specific public key belongs to a specific user or not. Public key is used so that anything signed with it can be decrypted by only its corresponding private key.
  \item A random set of peers (public keys and corresponding contact information such as IP address/port number) can be obtained without an adversary knowing who obtained a specific information. This assumption is important, otherwise, if one can determine who all obtained a specific peer ID from the sampling service, then it reduces the degree of anonymity.

      A random set of peers can readily be obtained using gossip based peer sampling \cite{jelasity}. We argue that peers who participate in the process of peer sampling for a relatively long time would encounter sufficiently large number of other peers to mitigate any set intersection analysis by an adversary.
  \item Proxies delegated by users of similar profiles can discover each other and carry out the aggregation task. Note that this last assumption is needed for the aggregation task, and is orthogonal to the anonymity issues. Gossip based mechanism like T-man \cite{Tman} or variants \cite{epiman} can be applied for this. Note also that the gossiping overheads for the various tasks like peer sampling and information aggregation can be amortized.
\end{enumerate}

We use the following notations while detailing the anonymous gossiping mechanism.
\small
\hrule
\begin{description}
  \item[$\alpha_{i}$] Public key and ID of peer $i$.
  \item[$E_{i}(.)$] Encryption of message with public key of peer $\alpha_{i}$, so that only she can decrypt it.
    \item[$\Phi_{i}$] Random set of peers that peer $\alpha_i$ has obtained somehow.
  \item[$\kappa_i$] A symmetric en/de-cryption key created by peer $\alpha_{i}$.
  \item[$\kappa_i(.)$] Message encrypted with the key $\kappa_i$.
  \item[$\pi_i$] Profile of peer $\alpha_i$. The records of the profile may/not need themselves to be perturbed or obfuscated for privacy preserving data analytics \cite{PPDM}. That is however an orthogonal issue.
  \item[$\pi'_i$] Aggregated/personalized clustered information corresponding to profile $\pi_i$.
\end{description}
\hrule
\normalsize
\section{Anonymous gossiping}
\label{sec:anongoss}
There are three logical phases for anonymous gossiping: (phase I) aggregation task delegation to a proxy in an anonymous manner, (phase II) the proxies carrying out the delegated aggregation tasks, and finally (phase III) obtaining the results back from the delegate in an anonymous manner.

The aggregation task (phase II) is an interesting problem on its own right. Either existing solutions \cite{semanticoverlay,swam,Gossple} or new ones may be applied for it. We consider it as a black-box and focus on the other two phases as described next.

In the description below we consider a scenario where $\Alice$ is using anonymous gossiping to carry out privacy preserved personalized aggregation.
\subsection{Delegation of aggregation task}
Aggregation task is delegated to a proxy as follows:

\begin{itemize}
  \item Obtain $\Phi_{\Alice}$, a moderately large and random subset of peers in the system, e.g., using peer-sampling \cite{jelasity}.
  \item Determine the candidate $\alpha_{\mu}$ for task delegation where $\alpha_{\mu} \in \Phi_{\Alice}$.

  $\Alice$ needs to send the message $msg = (\pi_{\Alice},\kappa_{\Alice})$ containing her profile and a symmetric key to this delegate anonymously.

Note that the message contains the profile, but not $\Alice$'s identity. However, if the profile itself contains identity revealing details (such as search terms from `ego search') then our approach can not provide anonymity. Generally speaking, possible obfuscation of the records using traditional PPDM techniques \cite{PPDM} may additionally be necessary.
  \item Choose $k$ other peers $\alpha_i \in \Phi_{\Alice}$, where $k$ is a random integer chosen uniformly from some predefined range, say [5, \ldots, 20]. These peers will be used to form an onion route \cite{Tor,onion} between $\Alice$ and the delegate $\alpha_{\mu}$.
  \item Send to peer $\alpha_{\rho_1}$ an onion encoded message \small $E_{\rho_1}(E_{\rho_2}(...(E_{\rho_{k-1}}(E_{\rho_k}(E_{\mu}(msg),\mu),\alpha_{\mu}),\alpha_{\rho_{k}}),...),\alpha_{\rho_2})$\\

\normalsize
When a peer $\alpha_{\rho_j}$ receives an onion encoded message from $\alpha_{\rho_{j-1}}$, she can only decrypt the outermost layer with her own private key. Upon decryption, it obtains another encrypted message along with the identity/address of the next node to which it should pass the same. Thus, intermediate nodes do not know how many nodes or who have already routed the message, nor the nodes who will subsequently do so. Each node only knows its immediate up \& down-stream peers for an onion route. Given that the nodes were chosen at random by the source further reduces chances of collusion. The random choice of the length $k$ of the onion route provides a further level of obfuscation and plausible deniability for $\Alice$. It also provides robustness against small scale opportunistic collusion among some of the intermediate nodes to unravel the route requester's identity. Onion routing is robust against traffic snooping provided a minimal amount of ambient traffic is present in the system \cite{onionattack}.
\end{itemize}
Once the designated delegate $\alpha_{\mu}$ obtains the $msg = (\pi_{\Alice},\kappa_{\Alice})$, it can carry out the Phase-II task of aggregation and analytics using $\pi_{\Alice}$ to compile $\pi'_{\Alice}$.
\subsection{Collecting aggregated information}
One option to collect the aggregated information is for $\Alice$ to probe the delegate, and pull the response along a (new) onion route. Note that $\Alice$ should not reveal her identity to the delegate, so the delegate can not know which node to send the response to. The response $\kappa_{\Alice}(\pi'_{\Alice})$ encrypted with the symmetric key originally sent by Alice, and digitally signed by the delegate $\alpha_{\mu}$ may be sent upstream along the onion route (since each node knows the immediate up \& down-stream neighbors of a route without the delegate having to know specifically the destination.

There are however some potential drawbacks with such a pull based approach. Firstly, since $\Alice$ does not know if and when the delegate has completed computation of $\pi'_{\Alice}$, she may have to initiate pull on multiple occasions. Secondly, and more fundamentally, a passive attacker can monitor the network traffic (sniffing for $\kappa_{\Alice}(\pi'_{\Alice})$) and detect the terminal point. This can be alleviated if the message is encrypted by the intermediate nodes at every hop using the public key of the immediate node upstream.

An alternative to pull is a blind gossip based push mechanism. Whenever proxy $\alpha_{\mu}$ needs to send the aggregated information $\pi'_{\Alice}$ back to $\Alice$, she can just flood the network with corresponding $\pi'_{\Alice}, \alpha_{\mu},\kappa_{\Alice}(\pi_{\Alice})$ digitally signed with its private key. On receiving any such flooded message originating from $\alpha_{\mu}$, $\Alice$ can still determine whether the message is indeed meant for her or not using $\kappa_{\Alice}(\pi_{\Alice})$ even if $\alpha_{\mu}$ is also proxy for other peers. $\Alice$ should continue forwarding the message in all cases, so that no one monitoring the network can identify her as the intended destination for the information. A possible optimization during flooding is that each peer propagates the message to only peers it has obtained onion routed messages within a past time window. That will ensure, in absence of churn, that the source ($\Alice$) gets the aggregated information, while avoiding a larger scale flooding.
\section{Concluding remarks}
\label{sec:conclude}
We have defined a new communication primitive, namely \emph{anonymous gossiping}, which can support privacy preserving data aggregation and analytics complementing traditional computational primitives for PPDA based on randomization or secure multi-party computation. We have provided a rough but concrete sketch of one way to realize anonymous gossiping by composing existing techniques like peer sampling and onion routing. Use of such mature techniques is expected to facilitate a quick implementation of anonymous gossiping.

Additionally, this paper opens interesting avenues spanning algorithms, implementation as well as analysis which forms our ongoing work.
Exploration of new, more efficient, robust and churn resilient algorithms for anonymous gossiping is one direction. Clever implementation, particularly amortizing the various gossiping overheads (needed during peer-sampling and aggregation) provide nice systems design opportunities. Threat analysis including quantifying the trade-offs between the degree of anonymity and the time and messaging overheads in the peer-sampling process is a third frontier.
\section*{Acknowledgements} This work was inspired from discussions with Anne-Marie Kermarrec in the context of the GOSSPLE project during a research visit by the author to the ASAP Inria research group pertaining to the said project. 
\bibliographystyle{plain}
\bibliography{AnonGoss}
\label{sec:References}

\end{document}